\input phyzzx
\sequentialequations
\overfullrule=0pt
\def\part{\not\hskip-3pt\partial}
\vglue-.50in
\font\bigbold=cmbx12 scaled\magstep2
\title{\bigbold AN INFINITE LIE ALGEBRA ASSOCIATED WITH 
\break  THE QUANTUM COULOMB FIELD}
\vglue-.25in
\author{Andrzej Staruszkiewicz\foot{Address after September 15, 1992 : 
Department of Theoretical Physics, Jagiellonian University, Krakow, Poland}}
\address{Department of Physics and Astronomy
\break University of South Carolina
\break Columbia, South Carolina 29208}
\vglue-.25in
\fourteenpoint
\abstract{The theory of the quantum Coulomb field associates with each Lorentz 
frame, i. e., with each unit, future oriented time-like vector $u$, the 
operator of the number of transveral infrared photons $N(u)$ and 
the phase $S(u)$ which is the coordinate canonically conjugated with 
the total charge $Q$: $[Q,S(u)]=ie$, $e$ being the elementary charge. 
It is shown that the operators $N(u)$, ${Q\over e}S(u)$ and 
$Q^2$ form an infinite Lie algebra. One can conclude from this algebra 
that $\Delta{N(u)}=(4/{\pi})Q^2$, where $\Delta$ is the Laplace 
operator in the Lobachevsky space of four-velocities $u$, thus 
relating the total charge $Q$ with the number of infrared photons.}
\vskip 2cm
\centerline{PACS numbers: 12.20.Ds}
\endpage

\par
A charged particle, when scattered, produces an infrared electromagnetic 
field which behaves at the spatial infinity like the inverse of distance. 
Gervais and Zwanziger [1] gave a clean way to separate the infrared field 
from the rest : performing the rescaling $\lim_{\lambda\rightarrow\infty}
{\lambda A_{\mu}(\lambda x)}$ one obtains an electromagnetic potential 
which is homogeneous of degree $-1$ and represents thus a pure infrared 
field free from high frequency contaminations. Now, from the physical 
point of view, the infrared electromagnetic field is a free dynamical 
system : scattered charges produce infrared fields but infrared fields 
do not scatter charges; there is no back reaction. This allows to treat 
the infrared field as a free field; such a treatment is physically 
rigorous. This means also that the infrared field can be quantized in the 
usual way [2] . The resulting field theory is remarkable as it contains 
the quantum analog of the classical Coulomb field 
and allows to make meaningful 
and meaningfully arrived at statements on the magnitude of the fine 
structure constant $e^2 /{\hbar c}$ . 
In particular the value $e^2/{\hbar c} = {\pi}$ is seen to be critical as 
it separates two kinematically distinct regimes 
of the quantum Coulomb field [3] . 
\par
In the quantum theory of infrared fields the classical Coulomb field 

$$
A_0 = {Q\over r}, {\bf A} = 0, \eqn\joe
$$
is a global solution of the classical equation of motion, to be used, 
for the sake of completeness, in the procedure of quantization. The theory is 
summarized in [2] on page 364 in the form of all relevant canonical 
commutation relations, written in a fixed Lorentz reference 
frame which we imagine as 
being at rest. We repeat this summary here for the reader's convenience: 

$$ 
[Q , S_0] = ie , [Q , c_{lm}] = 0 , [S_0 , c_{lm}] = 0 , 
\eqn\junia
$$

$$
[c_{lm} , c^{\dagger}_{l'm'}] = 4{\pi} e^2 \delta_{ll'} \delta_{mm'} . 
\eqn\jerry
$$
Here $Q$ is the total charge, $e$ is the elementary charge, $S_0$ 
is the phase, a coordinate canonically conjugated with the operator 
${Q\over e}$; $c^{\dagger}_{lm}$ and $c_{lm}$ are, respectively, creation 
and annihilation operators for transversal infrared photons; they are numbered 
by the numbers $l = 1, 2, ...$ and $m = -l, -l + 1, ... , l$ known from 
the theory of angular momentum. 
\par
The vacuum state $|0>$ is defined by the relations (Ref. [2], p. 364) 

$$ 
   c_{lm}|0> = 0 , <0|c^{\dagger}_{lm} = 0 , Q|0> = 0 , <0|Q = 0 , 
\eqn\jake
$$
which are standard for transversal photons but nonstandard for the total 
charge $Q$. The reason for this is that the total charge $Q$ is a Lorentz 
invariant quantity while the phase $S_0$ is not; hence any definition 
of the vacuum state involving the phase $S_0$ would necessarily violate 
the Lorentz invariance. 
\par
The operator 

$$ 
   N = {1\over {4 \pi e^2}} {\sum_{lm}} c^{\dagger}_{lm} c_{lm} 
\eqn\judy
$$
gives the number of transversal infrared photons; its spectrum consists of 
nonnegative integers, $N = 0 , 1 , 2 , ...$.

Now, the whole theory is Lorentz invariant as it may be represented as 
a quantum field theory of a massless scalar field ``living'' in the 
three-dimensional hyperboloid of unit space-like four-vectors. 
Therefore, all the previous relations may be written in exactly 
the same form in another Lorentz reference frame :

$$ 
   [Q , S'_0] = ie , 
N' = {1\over {4 \pi e^2}} {\sum_{lm}} c'^{\dagger}_{lm} c'_{lm} , 
\eqn\juli
$$
etc. 
\par
Prime over a quantity denotes the same quantity in the moving reference frame 
which, without loss of generality, will be assumed to move in the 
{\it z}-direction. 
There is no prime over $Q$ since $Q = Q'$ is a Lorentz scalar. 

We can formulate now the main goal of this Letter: to relate primed 
quantities $N'$, $S'_0$ with the nonprimed ones $N$, $S_0$. It will be seen 
that the appropriate relation has the form of an infinite Lie algebra. 

In what follows it will be convenient to simplify notation and to replace 
the pair of indices $(lm)$ by a single index $\alpha$. Thus the quantum 
mechanics of charge is simplified to the following form 

$$ 
[Q , S_0] = ie , [Q , c_{\alpha}] = 0 , [S_0 , c_{\alpha}] = 0 , 
\eqn\jima
$$

$$
[c_{\alpha} , c^{\dagger}_{\beta}] = 4 \pi e^2 \delta_{\alpha \beta} , 
\eqn\jimmy
$$
and 

$$ 
   c_{\alpha}|0> = 0 , <0|c^{\dagger}_{\alpha} = 0 , Q|0> = 0 , <0|Q = 0 . 
\eqn\jimbo
$$
\par
We shall prove two lemmas which together give the Lorentz transformation 
of the amplitudes $c_{\alpha}$ and $S_0$. 

{\it Lemma 1.}

The amplitude $c'_{\alpha}$ is a linear combination of the 
amplitudes $c_{\alpha}$ and $Q$ {\it but does not contain the phase} $S_0$

$$
   c'_{\alpha}=\sum_{\beta} A_{\alpha\beta} c_{\beta} + B_{\alpha} Q . 
\eqn\julian
$$
Moreover, the matrix $A_{\alpha \beta}$ is unitary. 
\par
Indeed, since $c'_{\alpha}|0> = 0$, $c'_{\alpha}$ must be a linear 
combination of those unprimed amplitudes which annihilate the vacuum state, 
i. e., $c_{\alpha}$ and $Q$. Unitarity of the matrix $A_{\alpha \beta}$ 
follows from the fact that the Lorentz transformation \julian\ must preserve 
the canonical commutation relations

$$
  [c'_{\alpha} , c'^{\dagger}_{\beta}] = 4 \pi e^2 \delta_{\alpha \beta} . 
\eqn\judea
$$
{\it Lemma 2.}
\par
The amplitude $S'_0$ is a linear combination of the amplitudes 
$S_0$, $c_{\alpha}$ and $c^{\dagger}_{\alpha}$ but {\it does not contain 
the charge} $Q$ 

$$
 S'_0 = S_0 - {1\over {4 \pi ie}}\sum_{\alpha\beta}\big(\overline B_{\alpha} 
A_{\alpha \beta} c_{\beta} - B_{\alpha} \overline A_{\alpha \beta} 
c^{\dagger}_{\beta}\big) . 
\eqn\jarre
$$
It is easy to check that this expression is indeed consistent with the Lorentz  
invariance of the commutation relations which involve the phase $S_0$: 
calculating $[Q , S'_0]$, $[c'_{\alpha} , S'_0]$ and $[c'^{\dagger} , S'_0]$ 
and using \julian\ one finds identity in each case. This does not show that 
the term proportional to the charge $Q$ on the right hand side is absent. 
Using, however, the explicit expression of $S'_0$ through the solutions 
of the wave equation on the three-dimensional hyperboloid of space-like 
unit four-vectors given in Ref. [2], one finds indeed that the phase 
$S'_0$ does not contain the total charge $Q$. 

To summarize, we have the following expressions for the Lorentz transformation 
of all the amplitudes : 

$$
c'_{\alpha} = \sum_{\beta} A_{\alpha\beta} c_{\beta} + B_{\alpha} Q , 
\eqn\juliet
$$

$$
c'^{\dagger}_{\alpha} = \sum_{\beta} \overline A_{\alpha\beta} 
c^{\dagger}_{\beta} + \overline B_{\alpha} Q , 
\eqn\joram
$$

$$
S'_0 = S_0 - {1\over {4\pi} ie} \sum_{\alpha\beta} \big(\overline B_{\alpha} 
A_{\alpha\beta} c_{\beta} - B_{\alpha} \overline A_{\alpha\beta} 
c^{\dagger}_{\beta} \big) , 
\eqn\jarek$$
where the matrix $A_{\alpha \beta}$ is unitary. 
We have 

$$
   N' = {1\over {4 \pi e^2}}\sum_{\alpha} c'^{\dagger}_{\alpha} c'_{\alpha} . 
\eqn\jenny
$$
Putting into \jenny\ $c'_{\alpha}$ and using unitarity of the matrix 
$A_{\alpha\beta}$ one obtains 

$$
  N' = N + {Q\over {4 \pi e^2}} \sum_{\alpha\beta} \big(\overline B_{\alpha} 
A_{\alpha\beta} c_{\beta} + B_{\alpha} \overline A_{\alpha\beta} 
c^{\dagger}_{\beta} \big) + {Q^2\over {4 \pi e^2}} \sum_{\alpha} 
|B_{\alpha}|^2 . 
\eqn\jannis
$$
It is shown in Ref. [2] that 

$$
  \sum_{\alpha} |B_{\alpha}|^2 = 8 e^2 (\lambda coth{\lambda} - 1) , 
\eqn\jan
$$
where $\lambda$ is the hyperbolic angle between the time axis of the 
moving frame and that of the rest frame. 

Let us take the commutator $[N , S'_0]$. Using \jarre\ and the obvious 
relations $[N , c^{\dagger}_{\alpha}] = c^{\dagger}_{\alpha}$, 
$[N , c_{\alpha}] = -c_{\alpha}$, one finds 

$$
  [N , S'_0] = {1\over {4 \pi ie}} \sum_{\alpha\beta} \big(\overline B_{\alpha} 
A_{\alpha\beta} c_{\beta} + B_{\alpha} \overline A_{\alpha\beta} 
c^{\dagger}_{\beta} \big) . 
\eqn\janus
$$
Hence 

$$
  N' = N + i {Q\over e} [N , S'_0] + 2{Q^2\over {\pi}} 
(\lambda coth{\lambda} - 1) . 
\eqn\joanna
$$
Since the total charge $Q$ commutes with the operator $N$, the last equation 
\joanna\ can be written in the form 

$$
[N , {Q\over e}S'_0] = i \big(N - N' + 2{Q^2\over {\pi}} 
(\lambda coth{\lambda} - 1)\big) . 
\eqn\joasia
$$
This equation has already the form characteristic for the Lie algebras. 
To close the algebra we need further $[N , N']$ and $[{Q\over e}S_0 , 
{Q\over e}S'_0]$. 

One has immediately from the previous formula \joasia\ and canonical 
commutation relations 

$$
 [N , N'] = i {Q\over e} (S'_0 - S_0) , [{Q\over e}S_0 , {Q\over e}S'_0] = 
i {Q\over e} (S_0 - S'_0) . 
\eqn\june
$$

\par
An objection may be raised that the operator $({Q\over e})S_0$ is 
``ill defined'' as it can be represented in the rest frame as 
the differential operator $(i{\partial \over \partial S_0} S_0)$, 
$S_0$ being an angular variable with the period $2\pi$. 
However, one sees that the right hand sides of all the commutators contain 
only {\it differences} of two phases which are perfectly well defined. Such 
differences may be also introduced in the left hand sides; for example, 
the commutator $[N , {Q\over e}S'_0]$ can be written as $[N , 
{Q\over e}(S'_0 - S_0)]$ since $[N , {Q\over e}S_0] = 0$. The same can be 
done for all other commutators, which shows that all the commutators written 
above have perfectly well defined meaning. 
\par
It will be useful to change notation. We introduce the unit, time-like, 
future oriented vector $u$ 
which indicates the time axis of the rest frame 
and a similar vector $v$ which indicates the time axis of 
the moving frame. All such vectors together form the Lobachevsky space of 
four-velocities. Using this notation we have 

$$
\big[N(u) , N(v) \big] = 
i {Q\over e} \big( - S(u) + S(v) \big) , 
\eqn\jasiu
$$

$$
\big[N(u) , {Q\over e}S(v) \big] = 
i \big(N(u) - N(v) + 
2{Q^2\over \pi}(uv) \big) , 
\eqn\jasio
$$

$$
  \big[{Q\over e}S(u) , {Q\over e}S(v) \big] = 
i {Q\over e} \big(S(u) - S(v) \big) , 
\eqn\jasko
$$
where $(uv) = {\lambda}coth{\lambda} - 1$ , $\lambda$ being 
the hyperbolic angle between $u$ and $v$ : 
$g_{\mu\nu}u^{\mu}v^{\nu} = cosh{\lambda}$. Commutators involving $Q^2$ 
are omitted because they are completely obvious.
\par
The Lobachevsky space of four-velocities is a three-dimensional 
locally Euclidean space of constant negative curvature with the metric 
induced by the metric of the four-dimensional space-time in which it is 
immersed. One can execute in this space all invariant operations of 
tensor analysis. Let us apply the Laplace operator $\Delta_v$ ($v$ indicates 
that it is to be applied at the point $v$) to both sides of the 
commutator

$$ 
\big[N(u) , {Q\over e}S(v)\big] = i \big(N(u) - N(v) 
+ 2{{Q^2}\over {\pi}}(uv)\big) . 
\eqn\jana
$$
One finds $\Delta_{v}(uv) = 2$ by direct calculation. 
Thus

$$\big[N(u) , {Q\over e} \Delta_{v}S(v)\big] = 
i \big(- \Delta_{v}N(v) + 4{Q^2\over {\pi}}\big) . 
\eqn\janula
$$
But $\Delta_{v}S(v) = 0$, as can be shown again from the explicit 
expression for $S(v)$ given in Ref. [2]. Hence, dropping the index 
$v$ which is not necessary anymore we have 

$$\Delta N = 4{Q^2\over {\pi}} . 
\eqn\jakosc
$$

This equation is remarkable as it connects the total charge $Q$ with 
the operator $N$ of the number of transversal infrared photons which has 
an absolute scale, its spectrum consisting of nonnegative integers. 
\par
This research was supported in part by the Polish Council for Scientific 
Research (Komitet Bada\'{n} Naukowych) and University of South Carolina. 
I thank Prof. Y. Aharonov, Prof. F. Avignone III and Prof. P. O. Mazur 
for hospitality. Stimulating discussions with Prof. Y. Aharonov and 
Prof. P. O. Mazur are gratefully acknowleged. 
\endpage

\centerline{\bf References}
\item {[1]}    J.-L. Gervais and D. Zwanziger, Phys. Lett. {\bf B94}, 389(1980)
\item {[2]}    A. Staruszkiewicz, Ann. Phys. (N.Y.) {\bf 190}, 354(1989).
\item {[3]}    A. Staruszkiewicz, Acta Phys. Pol. {\bf B23}, 591(1992).
\end